# Study of the Shadow of the Moon and Sun with VHE Cosmic Rays

M.O. Wascko[1]
[1]*Department of Physics, University of California, Riverside, Riverside, CA 92521, USA*
for the Milagro collaboration

## Abstract

Milagrito, a prototype for the Milagro detector, operated for 15 months in 1997-8 and collected $8.9 \times 10^9$ events. It was the first extensive air shower (EAS) array sensitive to showers intiated by primaries with energy below 1 TeV. The shadows of the sun and moon observed with cosmic rays can be used to study systematic pointing shifts and measure the angular resolution of EAS arrays. Below a few TeV, the paths of cosmic rays coming toward the earth are bent by the helio- and geo-magnetic fields. This is expected to distort and displace the shadows of the sun and the moon. The moon shadow, offset from the nominal (undeflected) position, has been observed with high statistical significance in Milagrito. This can be used to establish energy calibrations, as well as to search for the anti-matter content of the VHE cosmic ray flux. The shadow of the sun has also been observed with high significance.

## 1 Introduction:

Extensive air shower (EAS) arrays have been used to search for astrophysical point sources of ultra high energy (UHE) and very high energy (VHE) $\gamma$-rays for decades, from the PeV to EeV regions in the 70's, down to 10 TeV in the 90's. To distinguish a point source of $\gamma$-rays from the large isotropic background of cosmic ray protons and nuclei, a detector must have at least one of the following two capabilities: it must be capable of distinguishing between photon and hadron initiated showers, or it must have angular resolution sharp enough to detect a significant excess of events above the isotropic cosmic ray flux. Typically, hadron-initiated showers have higher muon content than photon initiated showers, but at lower energies this difference is not as striking. Thus in the VHE region, angular resolution is a critical parameter for detector performance, regardless of particle identification capabilities.

As they pass overhead during a transit, the moon and the sun block cosmic rays, so their shadows in the cosmic ray flux should be visible to EAS arrays with sufficiently good angular resolution (Clark, 1957). A detector with perfect angular resolution would see a sharp bucket shaped deficit of events of radius $0.26°$ centered at the expected position of the moon or sun. In reality, the deficit of events is spread out from the expected position due to finite angular resolution effects. It is thus possible to use the observed shadows of the moon and the sun in the cosmic ray flux to determine the angular resolution of an EAS detector (Alexandreas, et al. 1991). By comparing the observed position of the deficit to the expected position, the shadowing effect can be used to determine whether there are systematic pointing shifts (Alexandreas, et al. 1991).

In the TeV regime, the paths of charged cosmic rays are noticeably bent by the magnetic fields of the earth and the sun. Thus it is expected that the shadows of the moon and the sun will be offset from their nominal positions. The amount of the magnetic deflection varies with the rigidity, $\frac{|\vec{p}|}{Z}$, of the primary particle, as well as the magnitude and direction of the magnetic field. Cosmic rays approaching the earth from different directions sample different parts of the geomagnetic field, and the effects of the magnetic deflection on the moon's shadow will differ. The sun's shadow should be somewhat more dispersed than the moon's shadow due to the complexity and variability of the heliomagnetic field (Amenomori et al. 1993). The sun's magnetic field changes noticeably in magnitude on time scales of several years.

By studying the effects of the geomagnetic deflection on the moon shadow, especially as a function of incident angle, it is possible to gain some understanding of the energy response of the detector. Lower energy particles will be deflected more than higher energy particles, and this will smear the shape of the shadow in addition to deflecting its position. Energy resolution is traditionally one of the weaker aspects of EAS arrays, and thus the moon's shadow again offers itself as a useful tool for understanding the detector's capabilities. The effect of the magnetic field can also be used to search for the antiproton to proton ratio in the VHE cosmic ray flux, since negatively charged antiprotons would be bent in the opposite direction that positively charged protons would be bent (Urban et al. 1990).

## 2 Experimental Technique

Milagrito was the prototype stage of the Milagro Gamma Ray Observatory. Consisting of a single layer of upward facing photomultiplier tubes (PMTs) submerged beneath 1-2 meters of water, it operated between February 1997 and May 1998 and collected $8.9 \times 10^9$ air shower events (Atkins, et al., 1999). Simulations indicate that Milagrito is capable of detecting showers from primaries with energies as low as 100 GeV, with the median energy of detected showers varying as a function of zenith angle. The present analysis is based on the subset of events whose arrival direction was within $8^o$ of the moon's direction.

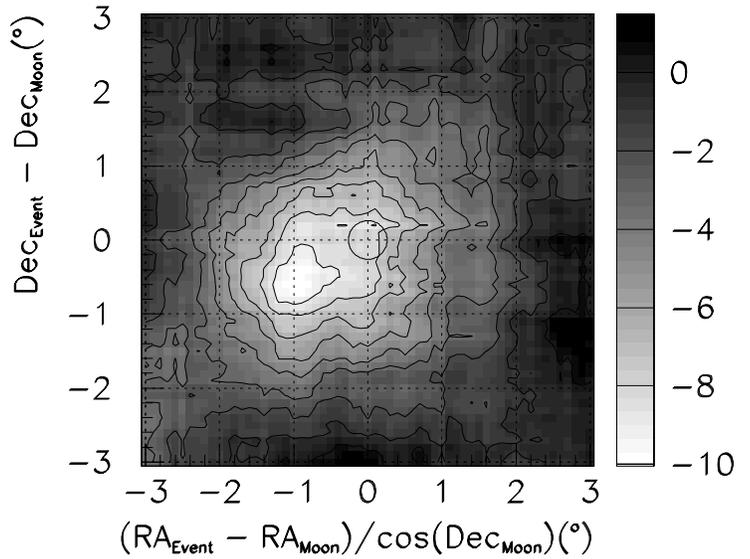

Figure 1: Two dimensional plot of the moon shadow. The background was calculated using the time-sloshing method (Alexandreas et al.,1993). The data and background maps were then smoothed with square bins $2.1^o$ on a side. The significance was calculated by the method of Li & Ma (Li & Ma, 1983). The circle shown is of radius $0.26^o$ and is centered at the undeflected position of the moon.

The methods used to extract the angular resolution and the systematic pointing shifts from the moon's shadow have been developed previously (Alexandreas, et al., 1991, Amenomori, et al., 1993). The event density as a function of the angular separation from the moon's position is calculated, and the shape of the deficit of events is analyzed using the maximum likelihood method. This assumes *a priori* knowledge of the shape of the resolution function; typically a two dimensional Gaussian (Alexandreas, et al., 1991), or a sum of two dimensional Gaussians (Amenomori, et al., 1993), is assumed. This analysis can yield both the most probable value for the width of the Gaussian point spread function, and the most probable position of the center of the deficit. The first result is the angular resolution of the detector, and the second gives the systematic pointing shift.

At TeV energies this is complicated by the fact that the position of maximum deficit is *expected* to be offset from the nominal position of the moon, due to the geomagnetic deflection. Thus a careful simulation study of the effects of the geomagnetic field is required. The same is true of the shadow of the sun.

## 3 Simulations

The Monte Carlo simulations of air showers and the detector are described elsewhere (Atkins, et al. 1999). A systematic pointing error in Milagrito was identified with the detector simulations. The effect is that air showers are reconstructed with zenith angles systematically closer to the horizon than the incident directions of the primary particles. This is thought to stem primarily from late light traveling laterally across the pond. The effect was removed from Milagro with the addition of reflective baffles on the PMTs, and the systematic error is not observed in the Milagro detector simulations. Simulations were also run in which cosmic ray primaries of varying rigidities were propagated through the Earth's magnetic field between the moon and the top of the atmosphere. The geomagnetic field was assumed to be a dipole, with the dipole axis coincident with the true magnetic poles. The angular deflection as a function of rigidity and incident direction was calculated in local detector coordinates and applied to events thrown from the moon as if it were a source of cosmic rays. These moon source events were then subtracted from a sample of separately simulated background events. This simulated moon shadow incorporates the effects of the systematic pointing error already identified with the Monte Carlo as well as the effect of the geomagnetic field. Further discrepancies between the data and the simulation would then indicate an additional systematic pointing shift.

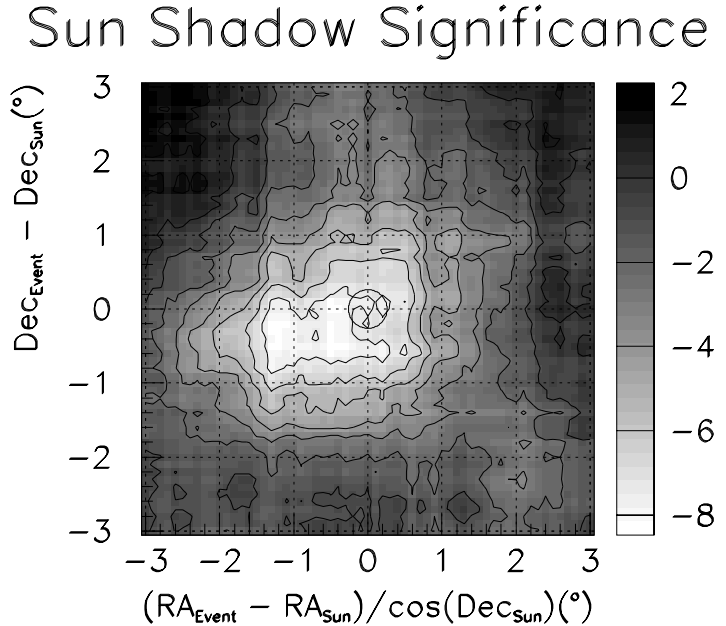

Figure 2: Two dimensional plot of the sun shadow. The background and significance were calculated in the same way as in the moon shadow plot.

## 4 Results and Conclusions

Barring systematic pointing shifts, one can make a prediction of the position of the center of the deficit of events, based on the simulations. Using this position as the location of the moon shadow, the event density as a function of angular separation from the moon shadow can be calculated. Unlike previous EAS moon shadow analyses, this calculation yields an upper limit of the angular resolution of Milagrito, rather than the resolution itself. This is because the spread of events in the deficit is due to the combined effects of the geomagnetic deflection and the finite angular resolution of Milagrito. To extract the angular resolution, one must use the simulations to unfold the two effects.

By carefully studying the shape of the moon shadow, one may be able to learn something about the energy response of the detector, since primaries of different rigidities will be deflected by different amounts. This may also lead to an energy calibration of Milagrito.

It is also possible to simulate a shadow of the moon in a flux of anti-matter cosmic rays. By combining such a simulation with the previous simulations, a prediction of the location and magnitude of a moon "anti-shadow", and then a search for such a feature in the data, can be made. In this way a measurement of the anti-matter content of the VHE cosmic ray flux can be made.

The shadow of the sun has also been observed with Milagrito, and is shown in Figure 2. Analyses of these measurements is in progress, and results will be presented.

## 5 Acknowledgments

This research was supported in part by the National Science Foundation, the U. S. Department of Energy Office of High Energy Physics, the U. S. Department of Energy Office of Nuclear Physics, Los Alamos National Laboratory, the University of California, and the Institute of Geophysics and Planetary Physics, The Reasearch Corporation, and CalSpace.

# References

Alexandreas, D.E. et al., 1991 Phys. Rev. D43, 1735.
Alexandreas, D.E. et al., 1993 Nucl. Intr. Meth. A328, 570.
Amenomori, M. et al., 1993 Phys. rev. D47, 2675.
Atkins et al. 1999, Nucl. Instr. Meth. A, in preparation.
Clark, G.W., 1957, Phys. Rev. 108, 450.
Li, T.P. & Ma, Y.Q., 1983 ApJ 272, 317.
Urban, M. et al., in *Astrophysics and Particle Physics*, Proceedings of the Topical Seminar, San Miniato, Italy, 1989, edited by G. Castellini et al. [Nucl. Phys. B(Proc. Suppl.), 14B, 223 (1990)].